\documentclass[10pt,journal,doublecolumn]{IEEEtran}
\usepackage{cite}
\usepackage{amsmath,amssymb,placeins}
\usepackage{threeparttable}
\usepackage{algorithmic}
\usepackage{graphicx}
\usepackage{textcomp}
\usepackage[mathscr]{euscript}
\usepackage{lipsum}
\usepackage{cuted}
\usepackage{mathtools}
\usepackage{multicol}
\usepackage{array}
\usepackage{makecell}
\usepackage{diagbox}
\usepackage[symbol]{footmisc}
\usepackage{caption}
\usepackage{tabularx}

\usepackage[mathscr]{euscript}
\usepackage[usenames]{color}
\usepackage{amsfonts}
\usepackage{latexsym}
\usepackage{subfigure}
\usepackage[format=hang]{subfig}
\usepackage{caption}
\usepackage{floatrow}
\usepackage{multirow}
\usepackage{epstopdf}
\newtheorem{theorem}{{{\textit{Theorem}}}}

\newtheorem{lemma}{{{\textit{Lemma}}}}

\newtheorem{definition}{{{\textit{Definition}}}}

\newtheorem{remark}{{{\textit{Remark}}}}

\newtheorem{example}{{{\textit{Example}}}}

\newcommand{\Mod}[1]{\ (\mathrm{mod}\ #1)}

\def\BibTeX{{\rm B\kern-.05em{\sc i\kern-.025em b}\kern-.08em
		T\kern-.1667em\lower.7ex\hbox{E}\kern-.125emX}}
		
		\def\@fnsymbol#1{\ensuremath{\ifcase#1\or *\or \dagger\or \ddagger\or
   \mathsection\or \mathparagraph\or \|\or **\or \dagger\dagger
   \or \ddagger\ddagger \else\@ctrerr\fi}}

\begin{document}
\title{A Direct Construction of Cross Z-Complementary Sets with Flexible Lengths and Large Zero Correlation Zone }
\author{Praveen Kumar, Sudhan Majhi, \IEEEmembership{Senior Member,~IEEE,}
       Subhabrata Paul  }

\IEEEpeerreviewmaketitle
\maketitle
\begin{abstract}
 This letter proposes a direct construction for cross Z-complementary sets (CZCSs) with flexible lengths and a large zero correlation zone (ZCZ). CZCS is an extension of the cross Z-complementary pair (CZCP). The maximum possible ZCZ width of a CZCP is half of its sequence length. In this letter, for the first time, a generalized Boolean function based construction of CZCSs with a large number of constituent sequences and a ZCZ ratio of $2/3$ is presented. 
For integers $m$ and $\delta$, the proposed construction produces CZCS with length expressed as $2^{m-1}+2^\delta$ ($0 \leq \delta <m-1,m\geq 4$), where both odd and even lengths CZCS can be obtained. 
Additionally, the constructed CZCS also feature a complementary set of the same length. 
 Finally, the proposed construction is compared with the existing works.
\end{abstract}
\section{Introduction}\label{sec:intro}
\IEEEPARstart{G}{olay} was the first to introduce Golay complementary pairs (GCPs) \cite{gol1961}. The aperiodic auto-correlation sum (AACS) of a GCP becomes zero for all shifts other than zero. Binary GCPs can only have lengths of the type $2^\alpha 10^\beta 26^\gamma$, where $\alpha, \beta,$ and $\gamma$ are all non-negative integers \cite{parker}. Using generalized Boolean functions (GBFs) over $\mathbb{Z}_{2^h}$, for any positive integer $h$, Davis and Jedwab provide a simple construction of GCP \cite{Davis}. The findings of Davis and Jedwab were further expanded by Paterson to include $q$-ary GCPs for even $q$ \cite{pater2000}. Recently, GCP of non-powers-of-two lengths using  GBFs is provided in \cite{praveen2}.  When the number of sequences is more than two it is called complementary set (CS) and its direct constructions are provided in \cite{pater2000,pmz2}. 
Fan \textit{et al.} introduced the concept of Z-complementary pairs (ZCPs) as a generalization of GCP \cite{fan2007}. Inside a certain zone known as the zero correlation zone (ZCZ), AACS of ZCP is zero. 
ZCPs, in contrast to GCPs, come in various ZCZ widths and arbitrary lengths \cite{zcp1,zcp2,zcp3,zcp4,zcp5,praveen1}.

Cross Z-complementary pair (CZCP) was produced as a consequence of Liu \textit{et al.} investigation into the training sequence design for the spatial modulation (SM) system \cite{czcp2}. AACS of CZCPs have both front-end and tail-end ZCZ, and aperiodic cross-correlation sum (ACCS) of CZCPs have only tail-end ZCZ. 
 For a perfect CZCP of length $N$, the largest ZCZ width is $N/2$, i.e., $Z_{max} = N/2$ \cite{czcp2}. 
 The length of perfect CZCPs is extremely constrained, though, binary perfect CZCPs exist with lengths $2^{\alpha+1}10^\beta26^\gamma$ where $\alpha+\beta +\gamma \geq 0$ \cite{czcp2}. 
The $q$-ary CZCPs with lengths of the type $10^\beta26^\gamma,2^{\alpha+1}10^\beta26^\gamma+2$, and $12\times 2^{\alpha+1}10^\beta26^\gamma $ have recently been proposed in \cite{czcp1,czcp5}. The majority of CZCP constructions are built on existing GCPs or Barker sequences \cite{czcp1,czcp2}. CZCPs can also be constructed \t{by using} existing ZCPs as suggested in \cite{czcp3,czcp4}. 
The direct constructions of CZCPs based on GBFs are also given in \cite{czcp1,czcp2,czcp4}. These CZCPs have lengths of the following types: $2^m ~(m\geq 2)$, or  $2^{m-1} + 2^{v+1} ~(0 \leq v \leq m-3,m\geq 4)$.
CZCPs exist only for even length and maximum ZCZ width is restricted only to half of the sequence length \cite{czcp2}. 
 
 To overcome these shortcomings of CZCPs, recently  Huang \textit{et al.} proposed many indirect constructions of the CZCSs having four sequences, acknowledging the constraints on the ZCZ width of the CZCPs \cite{czcs}. In \cite{czcp2}, identical CZCPs are combined to create cross Z-complementary sets (CZCSs) with the same length and same ZCZ ratio as CZCP. The construction of CZCSs containing a large number of constituent sequences based on GBF is proposed in \cite{praveen3}, but their ZCZ ratio is less than $1/2$.

In this letter, we present a GBF based construction of CZCSs of length of the form $2^{m-1}+2^\delta$, where $m\geq4$ and $0\leq \delta<m-1$. The proposed CZCSs have a large number of constituent sequences, flexible lengths, and large ZCZ widths. The proposed construction is capable of producing CZCSs of odd and even lengths with ZCZ ratios up to $2/3$. The proposed CZCSs have larger flock size as compared to \cite{czcs}, cover more length and have better ZCZ ratio than \cite{czcp2,praveen3}. Therefore the proposed direct construction is advantageous over \cite{czcp2,czcs,praveen3}. Moreover, the constructed CZCS also provides a CS of the same length.

The remainder of the letter is organised as follows. The preliminaries are covered in section II. The proposed CZCS construction based on GBF is covered in section III. Final conclusions are included in section IV.
\section{Preliminaries}
This section explains the fundamental definitions, notations, and previously known results required for the proposed construction.
\begin{definition}\label{def1} Let $\mathbf{u}$ $=(u_{0},u_{1}, \ldots, u_{N-1})$ and $\mathbf{v}$ $=(v_{0},v_{1}, \ldots, v_{N-1})$ be two $N$ length sequences over $\mathbb{Z}_{q}$. The aperiodic cross-correlation function (ACCF) is defined at a shift $\tau$ by
\begin{equation}\label{eqn 1}
\mathcal{C}\left({\mathbf{u}, \mathbf{v}}\right)(\tau)=\begin{cases}
\sum_{i=0}^{N-1-\tau} \omega^{u_{i}-v_{i+\tau}}, & 0 \leq \tau \leq N-1, \\
\sum_{i=0}^{N-1+\tau} \omega^{u_{i-\tau}-v_{i}}, & -N+1 \leq \tau \leq-1, \\
0, & |\tau| \geq N,
\end{cases}
\end{equation}
 where $q~(\geq 2)$ is an integer and $\omega=\exp(2\pi\sqrt{-1}/q)$. When $\mathbf{u}=\mathbf{v},~ \mathcal{C}(\mathbf{u}, \mathbf{v})(\tau)$ is known as aperiodic auto-correlation function (AACF) of $\mathbf{u}$ and is represented by $\mathcal{A}(\mathbf{u})(\tau)$.
 \end{definition}
\begin{definition}\label{def2} A pair of sequences $\mathbf{u}$ and $\mathbf{v}$ of length $N$, whose AACS is zero for all non-zero shifts inside a zone, constitutes a ZCP with ZCZ width $Z$, i.e.,
\begin{equation}\label{eqn2}
\mathcal{A}({\mathbf{u}})(\tau)+\mathcal{A}({\mathbf{v}})(\tau)=0, \text{ for all }  0 <\tau < Z.
\end{equation}
The pair $(\mathbf{u},\mathbf{v})$ is referred to as a GCP when $Z=N$.
\end{definition}
\begin{definition}\label{def3} 
 A set $A=\{\mathbf{a}_0,\mathbf{a}_1,\hdots,\mathbf{a}_{M-1}\}$, containing $M$ sequences each of length $N$ is known as an $(M,N,Z)$-CZCS if it satisfies the following two conditions
\begin{equation}\label{eqn3}
 \begin{aligned}
&{C1}: \sum_{i=0}^{M-1}\mathcal{A}({\mathbf{a}_i})(\tau)=0 \text {, for all }|\tau| \in \mathcal{T}_{1} \cup \mathcal{T}_{2};\\&
{C2}: \sum_{i=0}^{M-1}\mathcal{C}({\mathbf{a}_i, \mathbf{a}_{i+1}})(\tau)=0, \text{ for all } |\tau| \in \mathcal{T}_{2},
\end{aligned}  
\end{equation}
where $\mathbf{a}_M=\mathbf{a}_0$ and $\mathcal{T}_{1}=\{1,2, \ldots, Z\}$ and $\mathcal{T}_{2}=\{N-Z, N-Z+1, \ldots, N-1\}$. For an $(M,N,Z)$-CZCS, $M$ is the flock size, $N$ is the sequence length, and $Z$ is the ZCZ width. For the case when $M=2$, it is known as a CZCP.
\end{definition}

Corresponding to a GBF $f:\{0,1\}^m \rightarrow \mathbb{Z}_q$ of $m$ variables $x_1,x_2,\hdots,x_{m}$, the $\mathbb{Z}_q$-valued sequence is written as
$ \mathbf{f}=\left({f_0}, {f_1}, \hdots,{f_{2^m-1}}\right)$ ,
where $f_r=f({r_1},{r_2},\hdots,r_{m})$, and $r=({r_1},{r_2},\hdots,r_{m})$ is the 
binary vector representation of the integer $r$ \cite{pater2000}.
Corresponding to $m$ variables GBF $f$, the sequence $\mathbf{f}$ is of length $2^{m}$. The remainder of this letter concentrates on $q$-ary CZCSs of length in the form of non-power-of-two, where $q$ is a positive even integer. Here, the final $2^m-L$ components of the sequence $\mathbf{f}$ are removed to create the truncated $L$ length sequence $\mathbf{f}^{(L)}$ corresponding to the GBF $f$. For simplicity, when it is clear from the context, we will use $f$ to represent all three $(f,\mathbf{f},\mathbf{f}^{(L)})$. For a GBF $f$ in $m$ variables $x_1,x_2,\hdots,x_m$, $\bar{f}(x_1,x_2,\hdots,x_m)=f(\bar{x}_1,\bar{x}_2,\hdots,\bar{x}_m)$, where $\bar{x}_i=1-x_i, 1\leq i \leq m$.


\section{Proposed Construction}
This section describes a direct CZCS construction based on GBF with flexible lengths of the type $2^{m-1}+2^\delta$ ($0 \leq \delta<m-1,m\geq 4$).
\begin{theorem}\label{thm1}
For any positive integer $m \geq 4$ and even integer $q$, let $S_1,S_2,\hdots,S_k$ be a partition of the set $\{1,2,\hdots,m-1\}$, $1\leq k\leq m-1$. Let $S_\beta$ contains $m_\beta$ elements and $\pi_\beta$ be a bijective mapping from $\{1,2,\hdots,m_\beta\}$ to $S_\beta$ for $\beta=1,2,\hdots,k.$ Let us define a GBF $f:\{0,1\}^m \rightarrow \mathbb{Z}_q$ as
\begin{equation}\label{eqn4}
    f(x_1,x_2,\hdots,x_m)=\bar{x}_m\bar{g}+x_m g+\sum_{l=1}^{m-1}\lambda_lx_l +\lambda_0,
\end{equation}
where $\lambda_l \in \mathbb{Z}_q$, $0\leq l \leq m-1$ and $g:\{0,1\}^{m-1} \rightarrow \mathbb{Z}_q$ is defined as
\begin{equation}\label{eqn5}
    g(x_1,x_2,\hdots,x_{m-1})= \frac{q}{2} \sum_{\beta=1}^{k} \sum_{\gamma=1}^{m_{\beta}-1} x_{\pi_{\beta}(\gamma)} x_{\pi_{\beta}(\gamma+1)}.
\end{equation}
If sets $\{\pi_1(1),\pi_1(2),\hdots,\pi_1(\delta)\}$ and $\{1,2,\hdots,\delta\}$ are equal for a non-negative integer $\delta<m-1$, then the ordered set $A=\{\mathbf{a}_0,\mathbf{a}_1,\hdots,\mathbf{a}_{2^{k+1}-1}\}$ with the natural order induced by the vector $(\sigma_1,\sigma_2,\hdots,\sigma_{k+1})$ forms a $\left(2^{k+1},2^{m-1}+2^\delta,2^{\pi_{k}(1)-1}+2^\delta\right)$-CZCS by deleting the last $2^{m-1}-2^\delta$ elements of every sequence, where 
\begin{equation}\label{eqn6}
    \mathbf{a}_\sigma=f+\frac{q}{2}\left(\sum_{\beta=1}^{k}\sigma_\beta x_{\pi_\beta(1)}+\sigma_{k+1}x_m\right).
\end{equation}
\end{theorem}
\begin{IEEEproof}
The Appendix contains the proof.
\end{IEEEproof}
\begin{remark}\label{remark2}
The ZCZ width of the proposed CZCS is maximum when $\pi_k(1)=m-1$. For $\delta=m-2$ and $\pi_k(1)=m-1$, the $ZCZ_ {ratio}=\frac{Z}{N}=\frac{2^{m-2}+2^{m-2}}{2^{m-1}+2^{m-2}}=\frac{2}{3}$.
\end{remark}
Now, we provide an example to illustrate how the proposed construction is utilized to generate CZCSs with flexible length and large ZCZ width.
\begin{example}\label{ex1}
Let $m=5,q=4$, $S_1=\{1,2,3\}$ and $S_2=\{4\}$ such that $\pi_1(1)=1,\pi_1(2)=2,\pi_1(3)=3$ and $\pi_2(1)=4$. Let the GBF $f$ be given by $f=2\bar{x}_5\left(\bar{x}_1\bar{x}_2+\bar{x}_2\bar{x}_3\right)+2x_5\left(x_1x_2+x_2x_3\right)$. Consider the ordered set $$ 
\begin{aligned}
A=\{&f,f+2x_1,f+2x_4,f+2(x_1+x_4),f+2x_5\\&,f+2(x_1\!+x_5),f+2(x_4\!+x_5),f+2(x_1\!+x_4\!+x_5)\}.
\end{aligned}
$$
Since $\{\pi_1(1)\}=\{1\}$, by taking $\delta=1$ from \textit{Theorem} \ref{thm1} the set $A$ is a $2^4+2^1=18$ length CZCS with ZCZ width $Z=2^3+2^1=10$ and flock size $8$. So, the ZCZ ratio in this case is $10/18=5/9>1/2$. If we take $\delta=2$, then from \textit{Theorem} \ref{thm1}, $A$ becomes a $(8,20,12)$-CZCS, and the ZCZ ratio in this case is $3/5>1/2$. Also, since $\{\pi_1(1),\pi_1(2),\pi_1(3)\}=\{1,2,3\}$, so by taking $\delta=3$, the set $A$ is a $(8,24,16)$-CZCS from \textit{Theorem} \ref{thm1}. In this case, the CZCS achieves its maximum ZCZ ratio of $2/3$. For $\delta=3$, the AACF and ACCF values of 
sequences of the set $A$ at all the positive shifts including zero are listed in Table \ref{table 1} and Table \ref{table 2}, respectively. By setting $\delta=0$, we can get an odd length CZCS of length $2^4+2^0=17$ and ZCZ width $Z=2^3+2^0=9$, hence the ZCZ ratio in this case is $9/17>1/2$.
\end{example}
\begin{table}[h]
\centering
\caption{AACF of the sequences of set $A$ in Example \ref{ex1} for $\delta=3$.}
\resizebox{\textwidth}{!}{
\begin{tabular}{|l|l|}
\hline
       $\mathcal{A}(a_0)(\tau)$         &    (24,-1,2,13,0,7,2,5,8,-1,8,3,0,5,0,5,0,-1,6,-3,0,3,-2,1)                                                                                                                               \\ \hline
  $\mathcal{A}(a_1)(\tau)$         & (24,1,2,-13,0,-7,2,-5,8,1,8,-3,0,-5,0,-5,0,1,6,3,0,-3,-2,-1)                                     \\ \hline
 $\mathcal{A}(a_2)(\tau)$         & (24,-5,-2,5,0,-1,-2,1,-8,3,-4,3,0,-3,4,1,0,-1,6,-3,0,3,-2,1)
   \\ \hline
   $\mathcal{A}(a_3)(\tau)$         & (24,5,-2,-5,0,1,-2,-1,-8,-3,-4,-3,0,3,4,-1,0,1,6,3,0,-3,-2,-1)
    \\ \hline
 $\mathcal{A}(a_4)(\tau)$         & (24,-3,-2,7,0,5,-2,-1,8,-1,-8,3,0,-3,0,-3,0,1,-6,3,0,-3,2,-1)
    \\ \hline
 $\mathcal{A}(a_5)(\tau)$         & (24,3,-2,-7,0,-5,-2,1,8,1,-8,-3,0,3,0,3,0,-1,-6,-3,0,3,2,1)
    \\ \hline
 $\mathcal{A}(a_6)(\tau)$         & (24,-3,2,11,0,1,2,7,-8,-1,4,-9,0,1,-4,-3,0,1,-6,3,0,-3,2,-1)
    \\ \hline
 $\mathcal{A}(a_7)(\tau)$         & (24,3,2,-11,0,-1,2,-7,-8,1,4,9,0,-1,-4,3,0,-1,-6,-3,0,3,2,1)
    \\ \hline
$\sum_{\sigma=0}^7\mathcal{A}(a_\sigma)(\tau)$         & (192,0,0,0,0,0,0,0,0,0,0,0,0,0,0,0,0,0,0,0,0,0,0,0)
    \\ \hline
\end{tabular}}\label{table 1}
\end{table}
\begin{table}[h]
\centering
\caption{ACCF of the sequences of set $A$ in Example \ref{ex1} for $\delta=3$.}
\resizebox{\textwidth}{!}{
\begin{tabular}{|l|l|}
\hline
       $\mathcal{C}(a_0,a_1)(\tau)$         &    (0,1,0,-7,0,5,0,5,0,1,-2,-1,0,-1,0,-1,2,5,0,1,0,1,0,1,0,1)                                                                                                                               \\ \hline
  $\mathcal{C}(a_1,a_2)(\tau)$         & (0,-1,4,3,0,-9,-4,3,0,1,-2,3,0,3,2,-3,0,-1,0,-1,0,-1,0,-1)                                     \\ \hline
 $\mathcal{C}(a_2,a_3)(\tau)$         & (0,5,-4,1,0,13,4,1,0,-3,2,-1,0,-9,-2,1,0,1,0,1,0,1,0,1)
   \\ \hline
   $\mathcal{C}(a_3,a_4)(\tau)$         & (0,1,-4,-3,0,9,4,-3,0,1,2,-1,0,7,-2,-3,0,-1,0,-1,0,-1,0,-1)
    \\ \hline
 $\mathcal{C}(a_4,a_5)(\tau)$         & (0,3,0,-5,0,15,0,-1,0,1,-2,-1,0,7,2,-3,0,-1,0,-1,0,-1,0,-1)
    \\ \hline
 $\mathcal{C}(a_5,a_6)(\tau)$         & (0,1,4,5,0,1,-4,-3,0,-3,-2,-1,0,-9,2,1,0,1,0,1,0,1,0,1)
    \\ \hline
 $\mathcal{C}(a_6,a_7)(\tau)$         & (0,3,-4,-1,0,3,4,7,0,1,2,3,0,3,-2,-3,0,-1,0,-1,0,-1,0,-1)
    \\ \hline
 $\mathcal{C}(a_7,a_0)(\tau)$         & (0,-1,-4,-5,0,-1,4,3,0,1,2,-1,0,-1,-2,5,0,1,0,1,0,1,0,1)
    \\ \hline
$\sum_{\sigma=0}^7\mathcal{C}(a_\sigma,a_{\sigma+1})(\tau)$         & (0,12,-8,-12,0,36,8,12,0,0,0,0,0,0,0,0,0,0,0,0,0,0,0,0)
    \\ \hline
\end{tabular}}\label{table 2}
\end{table}
The proposed construction of CZCS is compared with existing work and given in Table \ref{table 3}. The best case scenarios have been considered in the comparison.
\begin{remark}
For $\delta=0$, the proposed construction can generate odd length CZCS.
\end{remark}
\begin{remark}
The set $A=\{\mathbf{a}_0,\mathbf{a}_1,\hdots,\mathbf{a}_{2^{k+1}-1}\}$ of \textit{Theorem} \ref{thm1} is a CS of length $2^{m-1}+2^\delta$, so the CSs of \cite{chen1,chen2} are the special cases of the proposed construction.  
\end{remark}
\begin{remark}
By taking $\delta=1$, we can get $2^{m-1}+2$ length CZCSs, which is also obtained in \cite{praveen3}. So CZCSs of \cite{praveen3} are the special case of the proposed construction.
\end{remark}
\begin{table}[h]
\centering
\caption{A comparison between the proposed and existing construction.}
\resizebox{\textwidth}{!}{
\begin{threeparttable}[t]
\begin{tabular}{|l|l|l|l|l|l|}
\hline 
         Ref.         & Method      & Flock size             & Length               & Constraint                &  ZCZ Ratio \tnote{1}                                                                                                                 \\ \hline
          \cite{czcp2}      & Indirect   & $M$           & $L$\tnote{*}       &$M\geq2$     & $ {{1}/{2}}$                                         \\ \hline
       \cite{czcs}         & Indirect & $4$          & $2^\alpha10^\beta26^\gamma$       &$\alpha,\beta,\gamma \in \mathbb{N}$      & $1$   \\ \hline
        \cite{praveen3}         & Direct  & $2^{n+1}$            &$2^{m-1}+2$             &$m\geq 4$,$n>0$              &$ >{1}/{4} $   \\ \hline
      
\textit{Theorem} \ref{thm1}   & Direct   & $2^{k+1}$       &$2^{m-1}+2^{\delta}$        &$m\geq 4,0 \leq \delta \leq m-1$    &  ${2}/{3}$                                 \\ \hline
\end{tabular}
\begin{tablenotes}
\item[1] maximum ZCZ ratio is considered.
\item[*] $L$ is the length of a GCP.
\end{tablenotes}
\end{threeparttable}}
\label{table 3}
\end{table}
\section{Conclusion}
We proposed a direct construction of CZCS with variable lengths and large ZCZ width. The proposed GBF based construction generates CZCSs with a ZCZ ratio upto $2/3$. The proposed CZCSs contain a large number of constituent sequences and generate both even and odd length sequences. Compared to the existing construction, the proposed construction is direct, generates CZCSs with larger ZCZ width and flexible lengths, and generalizes many existing works.
\section*{Appendix \\ Proof of Theorem 1}
Before we begin the proof of \textit{Theorem} \ref{thm1}, we introduce a few lemmas for proving our main theorem. We utilize the notations stated in \textit{Theorem} \ref{thm1} and define some basic notations as follows. Let the binary representations of two non-negative integers $r,s<2^m$ be $(r_1, r_2,\hdots,r_m)$ and $(s_1, s_2,\hdots,s_m)$, respectively.
\begin{lemma}\cite{chen1}\label{lem1}
If $r_i=s_i$ for $i=1,2,\hdots,\delta$, where $\delta$ is a positive integer, then $s\geq r+2^\delta$ for integers $r$ and $s$ with $0\leq r< s<2 ^m$ and $m\geq 2$.
\end{lemma}
\begin{lemma}\cite{chen1}\label{lem2}
For an integer $r$ with $2^{m-1} \leq r < 2^{m-1}+2^\delta$ where $1\leq \delta \leq m-1$ and $m \geq 2$. If the integer $r'$ has a binary vector representation that varies from $r$ solely at the  $\alpha$-th place for $\alpha \leq \delta$, then $2^{m-1} \leq r' <  2^{m-1}+2^\delta.$
\end{lemma}
\begin{lemma}\cite{shing}\label{lem3}
Using the same notations and  definitions as in \textit{Theorem} \ref{thm1}. If $r_{\pi_{\beta}(1)} \neq$ $s_{\pi_{\beta}(1)}$ for some $\beta \in\{1,2, \ldots, k\}$, then for any sequences $\mathbf{a}_\sigma=\left(a_{\sigma, 0}, a_{\sigma, 1}, \ldots, a_{\sigma, N-1}\right) \in A$, there exists $\mathbf{a}'_{\sigma}=$ $\left(a_{\sigma, 0}^{\prime}, a_{\sigma, 1}^{\prime}, \ldots, a_{\sigma, N-1}^{\prime}\right)=\mathbf{a}_{\sigma}+(q / 2) x_{\pi_{\beta(1)}} \in A$ such that $\omega^{a_{\sigma, r}-a_{\sigma, s}}+\omega^{a_{\sigma, r}^{\prime}+a_{\sigma, s}^{\prime}}=0$. Similarly, if $r_{m} \neq s_{m}$, then for any sequence $\mathbf{a}_{\sigma} \in A$, there exists $\mathbf{a}'_{\sigma}=\mathbf{a}_\sigma+(q / 2) {x}_{m} \in A$ such that $\omega^{a_{\sigma, r}-a_{\sigma, s}}+\omega^{a_{\sigma, r}^{\prime}+a_{\sigma, s}^{\prime}}=0$.
\end{lemma} 
\begin{lemma}\cite{shing}\label{lem4}
Suppose $r_{\pi_\beta(1)}=s_{\pi_\beta(1)}$ for $\beta=1,2, \ldots, k$. Let us consider three conditions as\\
  C1: $\hat{\beta}$ is the largest integer satisfying $r_{\pi_{\beta}(\gamma)}=s_{\pi_{\beta}(\gamma)}$ for $\beta=1,2, \ldots, \hat{\beta}-1$ and $\gamma=1, 2,\ldots, m_{\beta}$.\\
C2: $\hat{\gamma}$ is the smallest integer such that $r_{\pi_{\hat{\beta}}(\hat{\gamma})} \neq s_{\pi_{\hat{\beta}}(\hat{\gamma})}$.\\
C3: Let $r^{\prime}$ and $s^{\prime}$ be integers that differ from $r$ and $s$, respectively, at only one position $\pi_{\hat{\beta}}(\hat{\gamma}-1)$, i.e., $r_{\pi_{\hat{\beta}}(\hat{\gamma}-1)}^{\prime}=1-r_{\pi_{\hat{\beta}}(\hat{\gamma}-1)}$ and $s_{\pi_{\hat{\beta}}(\hat{\gamma}-1)}^{\prime}=1-s_{\pi_{\hat{\beta}}(\hat{\gamma}-1)}$.
If all the above conditions are satisfied, then we have 
\begin{equation}\label{eqn7}
    h_r-h_s-h_{r'}+h_{s'} \equiv \frac{q}{2} \Mod q,
\end{equation}
where
\begin{equation}\label{eqn8}
    h=g+\sum_{l=1}^{m-1}\lambda_lx_l+\lambda_0.
\end{equation}
\end{lemma}
\begin{IEEEproof}
[Proof of Theorem \ref{thm1}] In order to prove that the set $A$ constructed from \textit{Theorem} \ref{thm1} is a $(2^{k+1},2^{m-1}+2^\delta,2^{\pi_k(1)-1}+2^\delta)$-CZCS, we need to prove the two conditions of \textit{Definition} \ref{def3}, with $M=2^{k+1},N=2^{m-1}+2^\delta,$ and $Z=2^{\pi_k(1)-1}+2^\delta$.

First, we prove that the set $A$ is a CS, i.e., for $0<\tau<2^{m-1}+2^\delta$, the AACS of the sequences in set $A$ is zero, i.e.,
\begin{equation}\label{eqn9}
\sum_{\sigma=0}^{2^{k+1}-1}\mathcal{A}(\mathbf{a}_\sigma)(\tau)=\sum_{r=0}^{2^{m-1}-\tau-1}\sum_{\sigma=0}^{2^{k+1}-1}\omega^{\left(a_{\sigma,r}-a_{\sigma,r+\tau}\right)}=0 .
\end{equation}
Let $s = r + \tau$ for every integer $r$. 
Then we prove that (\ref{eqn9}) holds true in each of the three scenarios listed below.

\textit{Case I:} $r_{\pi_{\beta}(1)}\neq s_{\pi_\beta(1)}$ for some $1\leq \beta \leq k$, or  $r_m \neq s_m$.

Then from \textit{Lemma} \ref{lem3}, we get
\begin{equation}\label{eqn10}
   \sum_{\sigma=0}^{2^{k+1}-1}\omega^{\left(a_{\sigma,r}-a_{\sigma,s}\right)}=0.
\end{equation}

\textit{Case II:} $r_{\pi_{\beta}(1)}= s_{\pi_\beta(1)}$ for all $1\leq \beta \leq k$ and $r_m=s_m=0$.

There exists $\hat{\beta}$ and $\hat{\gamma}$ that meet the requirements (C1) and (C2) of \textit{Lemma} \ref{lem4}. Let $r'$ and $s'$ be numbers that satisfy condition (C3) of \textit{Lemma} \ref{lem4}. According to \textit{Lemma} \ref{lem4}, $s' = r' +\tau$ and $h_r-h_s-h_{r'}+h_{s'} \equiv q/2 \Mod q$. In this case, as $r_m=s_m=0$, so  $f_{r'}-f_r\equiv h_{r'}-h_r$ and $f_{s'}-f_s\equiv h_{s'}-h_s$ and hence $f_r-f_s-f_{r'}+f_{s'}\equiv q/2 \mod q$.
\begin{equation}\label{eqn11}
    a_{\sigma,r}-a_{\sigma,s}-a_{\sigma,r'}-a_{\sigma,s'}=f_r-f_s-f_{r'}+f_{s'}\equiv q/2 \Mod q.
\end{equation}
So, from the above equation, we get $\omega^{\left(a_{\sigma, r}-a_{\sigma, s}\right)}+\omega^{\left(a_{\sigma, r'}+a_{\sigma, s'}\right)}=0$.

\textit{Case III:} $r_{\pi_{\beta}(1)}= s_{\pi_\beta(1)}$ for all $1\leq \beta \leq k$ and $r_m=s_m=1$.

In this case, $2^{m-1} \leq r, s < 2^{m-1}+2^{\delta}$. We claim the existence of an integer $v \leq \delta$ such that $r_{\pi_{1}(v)} \neq s_{\pi_{1}(v)}$. On the contrary, let us assume that $v>\delta$. Since $\left\{\pi_{1}(1), \pi_{1}(2), \ldots, \pi_{1}(\delta)\right\}=\{1,2, \ldots, \delta\}$ we can write $r_{i}=s_{i}$ for $i=1,2, \ldots, \delta$, thus we can have $s \geq r+2^{\delta} \geq 2^{m-1}+$ $2^{\delta}$ according to \textit{Lemma} \ref{lem1}.
This contradicts the condition that $s < 2^{m-1}+2^{\delta}$. Now, we have $v \leq \delta$, and we let $v$ be the smallest integer such that $r_{\pi_{1}(v)} \neq s_{\pi_{1}(v)}$. Let $r^{\prime}$ and $s^{\prime}$ be integers that are different from $r$ and $s$, respectively only at one position $\pi_{1}(v-1)$, i.e., $r_{\pi_{1}(v-1)}^{\prime}=1-$ $r_{\pi_{1}(v-1)}$ and $s_{\pi_{1}(v-1)}^{\prime}=1-s_{\pi_{1}(v-1)}$; so $s^{\prime}=r^{\prime}+\tau$. Since $\left\{\pi_{1}(1), \pi_{1}(2), \ldots, \pi_{1}(\delta)\right\}=\{1,2, \ldots, \delta\}$ and $v-1<\delta$, we have $\pi_{1}(v-1) \leq \delta$. Therefore, according to \textit{Lemma} \ref{lem2}, 
$2^{m-1} \leq r^{\prime}, s^{\prime} < $ $2^{m-1}+2^{\delta}$. Following a similar approach as in \textit{Case III}, it is possible to conclude that there exists $r^{\prime}$ and $s^{\prime}$ such that $\omega^{\left(a_{\sigma, r}-a_{\sigma, s}\right)}+\omega^{\left(a_{\sigma, r'}+a_{\sigma, s'}\right)}=0$. 

Next, we prove that for $|\tau|\geq 
2^{m-1}-2^{\pi_k(1)-1}$, the ACCS of the sequences in set $A$ is zero, i.e.,
\begin{equation}\label{eqn12}
 \sum_{\sigma=0}^{2^{k+1}-1}\mathcal{C}({\mathbf{a}_\sigma, \mathbf{a}_{\sigma+1}})(\tau)\!=\!\sum_{r=0}^{2^{m-1}-\tau-1}\sum_{\sigma=0}^{2^{k+1}-1}\omega^{\left(a_{\sigma,r}-a_{\sigma+1,r+\tau}\right)}\!=\!0, 
\end{equation}
 where $\mathbf{a}_{2^{k+1}-1}=\mathbf{a}_0$.
 The proof of the same is split into three cases.

\textit{Case-I:} $r_{\pi_{k}(1)}\neq s_{\pi_k(1)}$. 

In this case, there exists $\mathbf{a}'_\sigma=\mathbf{a}_\sigma+(q/2)x_{\pi_{k}(1)} \in A$, i.e.,
$$
a'_{\sigma,r}=a_{\sigma,r}+\frac{q}{2}r_{\pi_{k}(1)},~a'_{\sigma+1,s}=a_{\sigma+1,s}+\frac{q}{2}s_{\pi_{k}(1)}.
$$
So, $a'_{\sigma,r}-a'_{\sigma+1,s}\equiv a_{\sigma,r}-a_{\sigma+1,s}+\frac{q}{2} \Mod q$ as $r_{\pi_{k}(1)}\neq s_{\pi_{k}(1)}$. Therefore, $\omega^{\left(a_{\sigma, r}-a_{\sigma+1, s}\right)}+\omega^{\left(a'_{\sigma, r}+a'_{\sigma+1, s}\right)}=0$.

\textit{Case-II:} $r_m \neq s_m$.

Similar to \textit{Case I}, there exist $\mathbf{a}'_\sigma=\mathbf{a}_\sigma+(q/2)x_m \in A$, such that $\omega^{\left(a_{\sigma, r}-a_{\sigma+1, s}\right)}+\omega^{\left(a'_{\sigma, r}+a'_{\sigma+1, s}\right)}=0$.

\textit{Case III:} $r_{\pi_{k}(1)} = s_{\pi_{k}(1)}$ 
and $r_m = s_m$.

In this case, the value of $\tau$ is
\begin{equation*}
  \begin{aligned}
\tau= s-r=\sum_{i=1}^{m}(s_i-r_i)2^{i-1} 
\leq 2^{m-1}-2^{\pi_{k}(1)-1}.
\end{aligned}  
\end{equation*}
So, $\tau=2^{m-1}-2^{\pi_{k}(1)-1}$, and so the binary vector representation of $r=(0,0,\hdots,r_{\pi_{k}(1)},0,\hdots,r_m)$ and $s=(1,1,\hdots,r_{\pi_{k}(1)},1,\hdots,r_m)$, i.e., all the components of $r$ are zero except possibly at $r_{\pi_{k}(1)},r_m$ and all the components of $s$ are $1$, except possibly at two positions where $r_{\pi_{k}(1)} = s_{\pi_{k}(1)}$ and
$r_m = s_m$. Let $(\sigma_1,\sigma_2,\hdots,\sigma_{k+1})$ and $(\sigma_1',\sigma_2',\hdots,\sigma_{k+1}')$ be the binary vector representation of integers $\sigma$ and $\sigma+1$, respectively. Now, depending upon the values of $r_{\pi_{k}(1)}$ and $r_m$, we have the following four sub-cases.

\textit{Sub-case I:} $r_{\pi_{k}(1)}=0$ and $r_m=0$.

In this case, $r=0$, which has binary representation as $(0,0,\hdots,0)$ and $s=r+\tau=2^{m-1}-2^{\pi_{k}(1)-1}$ whose binary vector representation is $(1,1,\hdots,0,1,\hdots,0)$. The difference between $r$-th and $s$-th term of the sequences $\mathbf{a}_\sigma$ and $\mathbf{a}_{\sigma+1}$, respectively is calculated as
\begin{equation}
    \begin{aligned}
    a_{\sigma,r}-a_{\sigma+1,s}&=\frac{q}{2}(m-k-1)+\lambda_0\\&-\left( \sum_{l=1}^{m-1}\lambda_l+\lambda_0+\frac{q}{2}\left(\sigma_1'+\sigma_2'+\cdots+\sigma_{k-1}'\right)\right)\\&=\lambda-\frac{q}{2}\left(\sigma_1'+\sigma_2'+\cdots+\sigma_{k-1}'\right),
    \end{aligned}
\end{equation}
where $\frac{q}{2}(m-k-1)-\sum_{l=1}^{m-1}\lambda_l=\lambda \in \mathbb{Z}_q$. The ACCS between $\mathbf{a}_\sigma$ and $\mathbf{a}_{\sigma+1}$ can be written as
\begin{equation}
\begin{aligned}
    \sum_{\sigma=0}^{2^{k+1}-1}\mathcal{C}({\mathbf{a}_\sigma, \mathbf{a}_{\sigma+1}})(\tau)&=\sum_{\sigma=0}^{2^{k+1}-1}\omega^{a_{\sigma,r}-a_{\sigma+1,s}}\\&=\omega^\lambda \sum_{\sigma=0}^{2^{k+1}-1}(-1)^{(\sigma_1'+\sigma_2'+\cdots+\sigma_{k-1}')}.
    \end{aligned}
\end{equation}
The above equation is zero as the sum  $(\sigma_1'+\sigma_2'+\cdots+\sigma_{k-1}')$ is $0$ as many times as it is $1$.

\textit{Sub-case II:} $r_{\pi_{k}(1)}=1$ and $r_m=0$.

In this case, the difference between $r$-th and $s$-th term of the sequences $\mathbf{a}_\sigma$ and $\mathbf{a}_{\sigma+1}$, respectively is given by

\begin{equation}
    \begin{aligned}
    a_{\sigma,r}-a_{\sigma+1,s}&=\frac{q}{2}(m-k-3)+\lambda_{\pi_{k}(1)}+\lambda_0+\frac{q}{2}\sigma_k\\&-\left( \sum_{l=1}^{m-1}\lambda_l+\lambda_0+\frac{q}{2}\left(\sigma_1'+\sigma_2'+\cdots+\sigma_{k}'\right)\right)\\&=\lambda'-\frac{q}{2}\left(\sigma_1'+\sigma_2'+\cdots+\sigma_{k}'-\sigma_k\right),
    \end{aligned}
\end{equation}
where $\frac{q}{2}(m-k-3)+\lambda_{\pi_{k}(1)}-\sum_{l=1}^{m-1}\lambda_l=\lambda' \in \mathbb{Z}_q$. So, similar to \textit{Sub-case I}, the ACCS
$$
\sum_{\sigma=0}^{2^{k+1}-1}C({\mathbf{a}_\sigma, \mathbf{a}_{\sigma+1}})(\tau)=0.
$$

\textit{Sub-case III:} $r_{\pi_{k}(1)}=0$ and $r_m=1$.

In this case, the difference is given by
\begin{equation}
    \begin{aligned}
    a_{\sigma,r}-a_{\sigma+1,s}&=\lambda_0+\frac{q}{2}\sigma_{k+1}-\lambda_0\\&-\left(\frac{q}{2}(m-k-3)+ \sum_{\substack{l=1 \\ l\neq \pi_k(1)}}^{m-1}\lambda_l+\frac{q}{2}\sum_{\substack{\beta=1 \\ \beta \neq \pi_k(1)}}^{k+1}\sigma_\beta'\right)\\&=\lambda''\!-\!\frac{q}{2}\left(\sigma_1'\!+\sigma_2'\!+\cdots+\!\sigma_{k+1}'\!-\!\sigma_{\pi_k(1)}'\!-\!\sigma_{k+1}\right),
    \end{aligned}
\end{equation}
where $-\frac{q}{2}(m-k-3)-\sum_{\substack{l=1 \\ l\neq \pi_k(1)}}^{m-1}\lambda_l=\lambda'' \in \mathbb{Z}_q$. So, in a similar manner to \textit{Sub-case I} the ACCS of the sequences is zero, i.e.,
$$
\sum_{\sigma=0}^{2^{k+1}-1}\mathcal{C}({\mathbf{a}_\sigma, \mathbf{a}_{\sigma+1}})(\tau)=0.
$$

\textit{Sub-case IV:} $r_{\pi_{k}(1)}=1$ and $r_m=1$.

In this case $r=2^{\pi_k(1)-1}+2^{m-1}$, and $s=r+\tau=2^{\pi_k(1)-1}+2^{m-1}+2^{m-1}-2^{\pi_k(1)-1}=2^m$, which is not possible as the sequence length $N= 2^{m-1}+2^\delta \leq 2^{m-1}+2^{m-2} <2^m$. So, this case is not feasible. So, the result follows.
\end{IEEEproof}
\bibliographystyle{IEEEtran}
\bibliography{reference}
\end{document}